\documentclass[12pt,showpacs,preprintnumbers,amsmath,amssymb,nofootinbib,APS]{revtex4}

\usepackage{dcolumn}
\usepackage{bm}
\usepackage[latin1]{inputenc}
\usepackage[spanish,english]{babel}
\usepackage{amsfonts}
\usepackage{amssymb}
\usepackage{graphicx}

\newcommand{\be}{\begin{equation}}
\newcommand{\ee}{\end{equation}}
\newcommand{\bea}{\begin{eqnarray}}
\newcommand{\eea}{\end{eqnarray}}

\begin{document}

\title{{\bf Computing Black Hole entropy in Loop Quantum Gravity  from a Conformal Field Theory perspective }}

\author{Iv\'an   \surname{Agull\'o}}
\email[]{Ivan.Agullo@uv.es} \affiliation{Enrico Fermi Institute and
Department of Physics, University of Chicago, Chicago, IL 60637
USA}\affiliation{Departamento de F\'{\i}sica Te\'orica and IFIC,
Centro Mixto Universidad de Valencia-CSIC. Facultad de F\'{\i}sica,
Universidad de Valencia, Burjassot-46100, Valencia, Spain}

\author{Enrique F.  \surname{Borja}}
\email[]{Enrique.Fernandez@uv.es} \affiliation{Departamento de
F\'{\i}sica Te\'orica and IFIC, Centro Mixto Universidad de
Valencia-CSIC. Facultad de F\'{\i}sica, Universidad de Valencia,
Burjassot-46100, Valencia, Spain}

\author{Jacobo  \surname{D\'{\i}az-Polo}}
\email[]{Jacobo.Diaz@uv.es}\affiliation{Institute for Gravitation
and the Cosmos, Physics Department, Penn State, University Park, PA
16802, U.S.A.} \affiliation{Departamento de Astronom\'{\i}a y
Astrof\'{\i}sica, Universidad de Valencia, Burjassot-46100,
Valencia, Spain}

\date{\today}

\begin{abstract}
Motivated by the analogy proposed by Witten between Chern-Simons and
Conformal Field Theories, we explore an alternative way of computing
the entropy of a black hole starting from the isolated horizon
framework in Loop Quantum Gravity. The consistency of the result
opens a window for the interplay between Conformal Field Theory and
the description of black holes in Loop Quantum Gravity.

\end{abstract}


\maketitle

\section{Introduction}
Ever since the pioneering work of Bekenstein \cite{Bekenstein74}
about the physical entropy of black holes, one of the main
challenges of quantum gravity has been to describe the microscopic
degrees of freedom responsible for this entropy. At the present time
there are several proposals that, in spite of their totally
different motivations, have reproduced the Bekenstein-Hawking law at
the leading order, furthermore showing an agreement in the first
order logarithmic correction. This proliferation of different
alternatives raises the puzzle of understanding the underlying
reason of this broad agreement. On the basis of the observation
that, in addition, most of these approaches involve Conformal Field
Theory (CFT) techniques at some stage, it has been suggested that
conformal symmetry could play a fundamental role in this scenario
(see \cite{Carlip} and references therein).

Loop Quantum Gravity (LQG) \cite{thiemann,rovelli} offers a detailed
description of the black hole horizon quantum states \cite{ABK}. In
the isolated horizon framework, a black hole is introduced as an
inner boundary of the spacetime manifold. Over this boundary,
constraints implementing the isolated horizon properties are
imposed. They reduce, already at the classical level, the $SU(2)$
gauge symmetry of the theory to a $U(1)$ gauge symmetry on the
horizon. These $U(1)$ degrees of freedom, that at the quantum level
fluctuate independently from the ones of the bulk, are described by
a Chern-Simons (CS) theory and are responsible for the horizon
entropy. By counting these CS degrees of freedom, a robustly
verified \cite{Meissner,PRL,ABBDV} linear behavior of entropy as a
function of the horizon area is obtained at the leading order,
showing in addition the existence of a first order logarithmic
correction.

On the other hand, E. Witten proposed \cite{Witten} the
correspondence between the Hilbert space of generally covariant
theories and the space of conformal blocks of a conformally
invariant theory. This idea was applied in \cite{KaulMajumdar} to
the computation of the entropy for a horizon described by a
$SU(2)$-CS theory, by putting its Hilbert space in correspondence
with the space of conformal blocks of a $SU(2)$-Wess-Zumino-Witten
(WZW) model.

The purpose of the present paper is to make use of Witten's
correspondence for the $U(1)$-CS theory describing the black hole
horizon in LQG, looking for some hints on the role of CFT techniques
in this framework. Taking into account the fact that this $U(1)$
group arises as the result  of a geometric symmetry breaking from
the $SU(2)$ symmetry in the bulk, one can still make use of the well
established correspondence between $SU(2)$ Chern-Simons and
Wess-Zumino-Witten theories. However, in this case it will be
necessary to impose restrictions on the $SU(2)$-WZW model in order
to implement the symmetry reduction.
Through this procedure we expect to eventually reproduce the counting of dimensions of the $U(1)$-CS Hilbert space.\\

\section{Black hole entropy counting}
Let us summarize the main features and results of the black hole
entropy counting in LQG in the isolated horizon framework
\cite{ABK}. On a space-like slice $\Sigma$, the geometry of the bulk
is described, as usual, by a spin network. Some of the spin network
edges, however, end at the horizon surface $S$ (the intersection of
$\Sigma$ and the isolated horizon), endowing it with an area given
by
\begin{equation}
A=8\pi\gamma \ell_P^2\sum_{I=1}^N\sqrt{j_I(j_I+1)}\ ,
\end{equation}
where $j_I\in\mathbb{N}/2$ label the $SU(2)$ irreducible
representations corresponding to the $N$ edges piercing the horizon,
$\gamma$ denotes the Barbero-Immirzi parameter and $\ell_P$ is the
Planck length. These edges carry an additional label
$m_I\in\{-j_I,-j_I+1,...,j_I\}$ (the corresponding spin projection)
characterizing their intersection with the horizon (punctures).

On the other hand, the horizon geometry is described by a
U(1)Chern-Simons theory defined over a sphere with $N$
distinguishable topological defects (corresponding with the
punctures).\footnote{The fact that  punctures are distinguishable is
related to the action of diffeomorphisms during the quantization
procedure \cite{ABK, thiemann}, and plays a key role in the entropy
counting.} The states of this theory are characterized by labels
$a_I\in\mathbb{Z}_{\kappa}$ (${\kappa}$ being the level of the CS
theory) quantifying the angle deficits that give rise to the
distributional curvature of the horizon concentrated at each
puncture. The spherical topology of the horizon implies that these
$a_I$ labels must satisfy the so called projection constraint
$\sum_I{a_I}=0$. The matching of both (bulk and horizon) geometries
through the boundary conditions gives rise to a relation between
$a_I$ and $m_I$ labels, that reads \be 2 m_I= - a_I\ \mod\  \kappa\ . \ee

For a given value $A$ of area, the entropy can be computed as
$S(A)=k_B \log{ \mathfrak{n}(A)}$, being $k_B$ the Boltzman constant
and $ \mathfrak{n}(A)$ the number of independent Chern-Simons states
compatible with the above constraints, taking into account the
distinguishable character of the punctures. This is to say, $
\mathfrak{n}(A)$ is the number of different $a_I$-labeled horizon
states (satisfying the projection constraint) such that, for each of
them, there exists (at least) one $(j_I, m_I)$-labeled piercing from
the bulk compatible with it and with the value $A$ of the horizon
area. The relation between $m_I$ and $a_I$ labels allows us then to
formulate the entropy counting as a well defined combinatorial
problem in terms only of the $m_I$ labels as in \cite{DL}. Then, $
\mathfrak{n}(A)$ can be rewritten as: $
\mathfrak{n}(A)=1+\sum_{A'\leq A} d(A')$, where $d(A)$ is the number
of all the finite, arbitrarily long, ordered sequences
$\vec{m}=(m_1,...,m_N)$ of non-zero half-integers, such that \be
\sum_{I=1}^N m_I=0\ , \ \ \sum_{I=1}^N \sqrt{|m_I|
(|m_I|+1)}=\frac{A}{8 \pi \gamma \ell_P^2}\ .\ee Explicit
expressions for the solution of this combinatorial problem were
obtained in \cite{ABBDV,rapid}. If we define $k_I=2|m_I|$ and the
occupancy numbers $n_k$ as the number of punctures carrying a label
value $m$ such that $k=2|m|$, then a set of numbers $\{n_k\}$,
$k=1,2,...$ characterizes a $\vec{m}$ sequence up to reorderings and
sign assignments for $m_I=\pm\frac{1}{2}k_I$. Thus, $d(A)$ can be
expressed in terms of the set $C$ of all the $\{n_k\}$ sets
compatible with a given area $A$ by associating two sources of
degeneracy to each of these sets $\{n_k\}$. The first is the number
$R(\{n_k\})$ of different ways of reordering the $k_I$ labels in
order to obtain all the corresponding ordered sequences
$\vec{k}=(k_1,...,k_N)$. The second source of degeneracy is the
number $P(\{n_k\})$ of different sing assignments for the associated
$m_I$ numbers, in such a way that the projection constraint is
satisfied. With this
\begin{equation}
d(A)= \sum_{\{n_k\}\in C}
R(\{n_k\})\times P(\{n_k\}),
\end{equation}
where the sum is extended over all the sets $\{n_k\}$ in $C$.\\
This set $C$ of all $\{n_k\}$ configurations compatible with a given
area eigenvalue can be computed analytically \cite{ABBDV} using
number-theory related techniques, through an exact characterization
of the horizon area spectrum of LQG. The factor $R(\{n_k\})$ has its
origin in the distinguishable character of punctures (acquired in
the process of quantization of geometry) and can be obtained from
basic combinatorics as $R(\{n_k\})= (\sum_k{n_k})!/\prod_k{n_k!}$,
where the sum and product are extended to all values of $k$ (note
that, in practice, for a finite value $A$ of area all the sums and
products are always finite). Finally, the factor $P(\{n_k\})$
accounts for the dimensionality of the Hilbert space of the U(1)-CS
theory once the boundary conditions have been fixed and was obtained
in \cite{ABBDV,rapid} to be:

\be
\label{CS} P(\{n_k\})=\frac{1}{2 \pi} \int_0^{2 \pi} d\theta
\prod_k{n_k 2 \cos{(k \theta)}}\ .
\ee

\section{Implementing the analogy between Chern-Simons and Wess-Zumino-Witten.}
Let us begin by recalling the classical scenario and how the
symmetry reduction takes place at this level. The geometry of the
bulk is described by a $SU(2)$ connection, whose restriction to the
horizon $S$  gives rise to a $SU(2)$ connection over this surface.
As a consequence of imposing the isolated horizon boundary
conditions this connection is reduced to a $U(1)$ connection. In
\cite{ABK} this reduction is carried out, at the classical level,
just by fixing a unit vector $\vec{r}$ at each point of the horizon.
By defining a smooth function $r:S\to su(2)$ a $U(1)$ sub-bundle is
picked out from the $SU(2)$ bundle. This kind of reduction can be
described in more general terms as follows (see, for instance,
\cite{bojowald}). Let $P(SU(2),S)$ be a $SU(2)$ principal bundle
over the horizon, and $\omega$ the corresponding connection over it.
A homomorphism $\lambda$ between the closed subgroup $U(1)\subset
SU(2)$ and $SU(2)$ induces a bundle reduction form $P(SU(2),S)$ to
$Q(U(1),S)$, being $Q$ the resulting $U(1)$ principal bundle
with reduced $U(1)$ connection $\omega'$.
This $\omega'$ is obtained, in this case, from the restriction of $\omega$ to $U(1)$.\\
All the conjugacy classes of homomorphisms $\lambda: U(1) \to SU(2)$
are represented in the set $Hom(U(1), T(SU(2)))$, where $T(SU(2))=\{
diag(z,z^{-1})|z=e^{i \theta}\in U(1)\}$ is the maximal torus of
$SU(2)$. The homomorphisms in $Hom(U(1),T(SU(2)))$ can be
characterized by \be \lambda_p:z\mapsto diag(z^p,z^{-p})\ ,\ee for
any $p\in\mathbb{Z}$. However the generator of the Weyl group of
$SU(2)$ acts on $T(SU(2))$ by $diag(z,z^{-1})\mapsto
diag(z^{-1},z)$. If we divide out by the action of the Weyl group we
are just left with those maps $\lambda_p$ with $p$ a non-negative
integer, $p\in\mathbb{N}_0$, as representatives of all conjugacy
classes. These $\lambda_p$ characterize then all the possible ways
to carry out the symmetry breaking from the $SU(2)$ to the $U(1)$
connection that will be quantized later.

However, one can follow the alternative approach of first quantizing
the $SU(2)$ connection on $S$ and imposing the boundary conditions
later on, at the quantum level. This would give rise to a $SU(2)$-CS
theory on the horizon to which the boundary conditions have to be
imposed. The correspondence with conformal field theories can be
used at this point to compute the dimension of the Hilbert space of
the $SU(2)$-CS as the number of conformal blocks of the $SU(2)$-WZW
model, as it was done in \cite{KaulMajumdar}. It is necessary to
require, then, additional restrictions to the $SU(2)$-WZW model that
account for the symmetry breaking, and consider only  the degrees of
freedom corresponding to a $U(1)$ subgroup.

Let us briefly review the computation in the $SU(2)$ case, to later
introduce the symmetry reduction. The number of conformal blocks of
the $SU(2)$-WZW model\footnote{Notice that, though we are omitting
the $\kappa$ subindex, the group of the WZW theory is in fact the
quantum group $SU(2)_{\kappa}$. The $\kappa$ dependence is implicit
in the allowed sets of representations $\mathcal{P}$.}, given a set
of representations $\mathcal{P}=\{j_1,j_2,...,j_N\}$, can be
computed in terms of the so-called fusion numbers $N_{il}^r$
\cite{difrancesco} as \be N^{\mathcal{P}}=\sum_{r_i} N_{j_1
j_2}^{r_1} N_{r_1 j_3}^{r_2}...N_{r_{N-2} j_{N-1}}^{j_N}\ .\ee These
$N_{il}^r$ are the number of independent couplings between three
primary fields, {\it i.e.} the multiplicity of the $r$-irreducible
representation in the decomposition of the tensor product of the $i$
and $l$ representations $[j_i]\otimes[j_l]=\bigoplus_r N_{il}^r
[j_r]$. This expression is known as a fusion rule. $N^{\mathcal{P}}$
is then the multiplicity of the $SU(2)$ gauge invariant
representation ($j=0$) in the decomposition of the tensor product
$\bigotimes_{i=1}^N [j_i]$ of the representations in $\mathcal{P}$ .
The usual way of computing $N^{\mathcal{P}}$ is using the Verlinde
formula \cite{difrancesco} to obtain the fusion numbers. But
alternatively one can make use of the fact that the characters of
the $SU(2)$ irreducible representations,
$\chi_i=\sin{[(j_i+1)\theta]}/\sin{\theta}$, satisfy the fusion
rules $\chi_i \chi_j=\sum_r N_{ij}^r \chi_r$. Taking into account
that the characters form an orthonormal set with respect to the
$SU(2)$ scalar product, $\langle \chi_i|\chi_j\rangle =\delta_{i
j}$, one can obtain the number of conformal blocks just by
projecting the product of characters over the gauge invariant
representation \be N^{\mathcal{P}}=\langle
\chi_{j_1}...\chi_{j_N}|\chi_0\rangle= \int_0^{2\pi}
\frac{d\theta}{\pi} \sin^2{\theta} \prod_{I=1}^N
\frac{\sin{[(j_I+1)\theta]}}{\sin{\theta}}\ . \ee This expression is
equivalent to the one obtained in \cite{KaulMajumdar} using the
Verlinde formula; it gives rise to the same result for every set of
punctures $\mathcal{P}$.

To implement, now, the symmetry breaking we have to restrict the
representations in $\mathcal{P}$ to a set of $U(1)$ representations.
This corresponds in the case of Chern-Simons theory to performing a
symmetry reduction locally at each puncture. It is known that each
$SU(2)$ irreducible representation $j$ contains  the direct sum of
$2j+1$ $U(1)$ representations $e^{i j \theta} \oplus e^{i (j-1)
\theta}\oplus...\oplus e^{-i j \theta}$. One can make an explicit
symmetry reduction by just choosing one of the possible restrictions
of $SU(2)$ to $U(1)$ which, as we saw above, are given by the
homomorphisms $\lambda_p$. This corresponds here to pick out a
$U(1)$ representation of the form $e^{i p \theta} \oplus e^{-i p
\theta}$ with some $p\leq j$. The fact that we will be using these
reducible representations, consisting of $SU(2)$ elements as $U(1)$
representatives, can be seen as a reminiscence from the fact that
the $U(1)$ freedom has its origin in the reduction from $SU(2)$.

Having implemented the symmetry reduction, let us compute the number
of independent couplings in this $U(1)$-reduced case. Of course, we
are considering now $U(1)$ invariant couplings, so we have to
compute the multiplicity of the $m=0$ irreducible $U(1)$
representation in the direct sum decomposition of the tensor product
of the representations involved. As in the previous case, this can
be done by using the characters of the representations and the
fusion rules they satisfy. These characters can be expressed as
$\tilde{\eta}_{p_I}= e^{i p_I \theta}+e^{-i p_I \theta}= 2 \cos{p_I
\theta}$. Again, we can make use of the fact that the characters
$\eta_i$ of the $U(1)$ irreducible representations are orthonormal
with respect to the standard scalar product in the circle. Then, the
number we are looking for is given by

\be \label{WZW} N^{\mathcal{P}}_{U(1)}=\langle
\tilde{\eta}_{p_1}...\tilde{\eta}_{p_N}|\eta_0\rangle=\frac{1}{2\pi}
\int_0^{2 \pi} d\theta \prod_I^N 2 \cos{p_I \theta}\ ,\ee
where $\eta_0=1$ is the character of the $U(1)$ gauge invariant irreducible representation.
We can see that this result is exactly the same as the one obtained for $P(\{n_k\})$ in Eq. (\ref{CS}), coming from the $U(1)$-CS theory, just by identifying the $p_I$ with $k_I$ labels.\\

\section{Remarks and Conclusions}
Let us put this result in context with the entropy counting. As
explained above, in computing the entropy of a black hole within
LQG, there are several contributions involved. Some of them are
related with the LQG framework, like the computation of $C$ by
characterizing the black hole area spectrum, or $R(\{n_k\})$ due to
the distinguishability of the punctures originated in the
quantization process. The term $P(\{n_k\})$, however, is related
with the CS theory on the horizon. Once one has introduced all the
conditions imposed by the LQG framework, what is left is the
counting of states of a CS theory subject to some external inputs.
If there is any connection between this CS theory and a Conformal
Field Theory, one should expect this CFT to reproduce precisely this
term $P(\{n_k\})$, subject to the same external inputs. This is
exactly what we observe here by identifying the $p_I$ and $k_I$
labels. We are, thus, proposing a precise implementation of Witten's
analogy through this symmetry reduced counting that yields the
expected result.

From the physical point of view, the main change we are introducing,
besides using the CS-CFT analogy, is to impose the isolated horizon
boundary conditions at the quantum level, instead of doing it prior
to the quantization process. This is a preliminary step in the
direction of introducing a quantum definition of isolated horizons.

It is very interesting to observe that, as shown in \cite{largo},
the contribution to entropy of $P(\{n_k\})$ has a linear growth with
area, including a logarithmic correction. The $R(\{n_k\})$ term, on
the other hand, introduces very particular quantum effects, that are
specific from LQG. In particular, the stair-like behavior appearing
at the Planck scale in the entropy-area relation \cite{PRL,ABBDV},
has its origin in this factor. Thus, the picture obtained here seems
to be compatible with the proposed role of conformal symmetry on the
first order behavior of black hole entropy.

\section*{Acknowledgment}

We thank V. Aldaya, A. Ashtekar, J. A. de Azcarraga, R. Coquereaux
and L. Freidel for interesting discussions and suggestions. We
specially thank F. Barbero and E. Villasenor for many discussions,
strong ecouragement, and a careful reading of the manuscript. I. A.
thanks R. Wald for his kind hospitality at the University of Chicago
where part of this work was done. J. D. thanks A. Ashtekar  for his
kind hospitality at the IGC at Penn State during the realization of
part of this work.

This work was in part supported by Spanish grants
FIS2008-06078-C03-02, FIS2008-01980 and ESP2005-07714-C03-0, the
NSF grant PHY0854743 and the Eberly research founds of Penn State, and the NSF grant PHY04-56619 to the University of Chicago.
I. A. and J. D. acknowledge financial support provided be the
Spanish Ministry of Science and Education under the FPU program.


\begin{thebibliography}{99}


\bibitem{Bekenstein74} J.D.~Bekenstein, Phys. Rev. D {\bf 9}, 3292-3300
(1974).

\bibitem{Carlip} S.~Carlip, Phys. Rev. Lett. {\bf 82}, 2828 (1999); Phys. Rev. Lett. {\bf 99}, 021301 (2007);  Gen. Rel. Grav. {\bf 39}, 1519 (2007).

\bibitem{thiemann} T.~Thiemann. \emph{Modern canonical quantum general relativity}, CUP, Cambridge (2007).

\bibitem{rovelli} C.~Rovelli. \emph{Quantum Gravity}, CUP, Cambridge (2004).



\bibitem{ABK} A.~Ashtekar, J.~Baez, A.~Corichi, K.~Krasnov, Phys. Rev. Lett. {\bf 80}, 904 (1998); A.~Ashtekar, J.~Baez, K.~Krasnov, Adv. Theor. Math. Phys. {\bf 4}, 1 (2000).
\bibitem{Meissner} K. A.~Meissner, Class. Quantum Grav. {\bf 21}, 5245 (2004).
\bibitem{PRL} A.~Corichi, J.~D\'\i az-Polo, E.~Fern\'andez-Borja, Class. Quantum Grav. {\bf 24}, 243 (2007); Phys. Rev. Lett. {\bf 98}, 181301 (2007).
\bibitem{ABBDV}
I. Agullo, J.F. Barbero G., J. Diaz-Polo, E. F. Borja, E. J. S.
Villaseñor, Phys. Rev. Lett. {\bf 100}, 211301 (2008).
\bibitem{Witten} E.~Witten, Commun. Math. Phys {\bf 121}, 351 (1989).
\bibitem{KaulMajumdar} R.~Kaul and P.~Majumdar, Phys. Lett. B {\bf 439}, 267
(1998); Phys. Rev. Lett. {\bf 84} 5255 (2000).



\bibitem{DL} \label{DL} M.~Domagala, J.~Lewandowski, Class. Quantum Grav. {\bf 21}, 5233 (2004).


\bibitem{rapid} J.F.~Barbero G., E. J. S.~Villaseñor, Phys. Rev. D, {\bf 77}, 121502(R) (2008).


\bibitem{bojowald} M.~Bojowald, H.A..~Kastrup, Class. Quantum Grav. {\bf 17}, 3009 (2000).

\bibitem{difrancesco} P.~Di Francesco, P.~Mathieu and D.~Senechal \emph{Conformal Field
Theory}. Springer, New York(1997).


\bibitem{largo}
I. Agullo, J.F. Barbero G., J. Diaz-Polo, E. F. Borja, E. J. S.
Villaseñor, {\it in preparation}.


\end{thebibliography}
\end{document}